\documentclass[preprint,onecolumn,a4paper,superscriptaddress,floatfix,11pt,nolongbibliography]{revtex4-1}

\usepackage[T1]{fontenc}
\usepackage{graphicx}
\usepackage{dcolumn}
\usepackage{bm}
\usepackage{color}
\usepackage{amsmath}
\usepackage{amssymb}
\usepackage{hyperref}
\usepackage{subfigure}
\usepackage{float}
\usepackage{bigints}
\usepackage{changes}
\usepackage{array}
\usepackage{tabularx}

\newcommand{\rp}{R_p}
\newcommand{\SF}{FCNN SF}

\begin{document}
\title{Machine Learning Scoring Functions for Drug Discovery from Experimental and Computer-Generated Protein-Ligand Structures: Towards Per-Target Scoring Functions}

\author{F. Pellicani}
\affiliation{School of Science and Technology, Physics Division, University of Camerino, I-62032 Camerino (MC), Italy}
\author{D. Dal Ben}
\affiliation{School of Pharmacy, Medicinal Chemistry Unit, University of Camerino, I-62032 Camerino (MC), Italy}
\author{A. Perali}
\affiliation{School of Pharmacy, Physics Unit, University of Camerino, Via Madonna delle Carceri 9, I-62032 Camerino (MC), Italy}
\author{S. Pilati}
\affiliation{School of Science and Technology, Physics Division, University of Camerino, I-62032 Camerino (MC), Italy}
\affiliation{INFN-Sezione di Perugia, I-06123 Perugia, Italy}

\begin{abstract}
In recent years, machine learning has been proposed as a promising strategy to build accurate scoring functions for computational docking finalized to numerically empowered drug discovery. However, the latest studies have suggested that over-optimistic results had been reported due to the correlations present in the experimental databases used for training and testing. Here, we investigate the performance of an artificial neural network in binding affinity predictions, comparing results obtained using both experimental protein--ligand structures as well as larger sets of computer-generated structures created using commercial software. Interestingly, similar performances are obtained on both databases. We find a noticeable performance suppression when moving from random horizontal tests to vertical tests performed on target proteins not included in the training data. The possibility to train the network on relatively easily created computer-generated databases leads us to explore per-target scoring functions, trained and tested ad-hoc on complexes including only one target protein. Encouraging results are obtained, depending on the type of protein being~addressed. 
\end{abstract}

\maketitle

\section{Introduction}

Scoring functions (SFs) are popular models often adopted by medicinal chemists to perform computational docking and to score poses of candidate ligands in the pockets of target proteins~\cite{kulharia2008information,jain2006scoring}. They are expected to attribute the best score to the correct pose, to accurately predict the binding affinity (or a score proportional to the affinity), and to prioritize active ligands over inactive ones. The latter task, in particular, plays a pivotal role in virtual screening~\cite{walters1998virtual}. It allows for identifying the most promising drug molecules, thus reducing the time and the cost otherwise spent for in vitro screening~\cite{wienkers2005predicting}. 
Due to their importance in drug discovery~\cite{drews2000drug}, the development of SFs has been the focus of intense research endeavors for decades. Modern SFs are often grouped~\cite{liu2015classification} into empirical~\cite{gohlke2001statistical}, knowledge based~\cite{gohlke2000knowledge}, and force fields~\cite{yin2008medusascore} SFs. More recently, SFs have been implemented also using data-based approaches via machine-learning (ML) techniques~\cite{ain2015machine,li2020machine,li2021machine}. This strategy is favored by the increasing amount of crystallographic protein--ligand structures~\cite{palmer2003x}, and it aims at exploiting the effectiveness of ML techniques in extracting useful information from big databases. An influential article was published in 2010~\cite{Ballester2010}. Therein, the authors adopted a relatively simple ML model, namely, random forest regression~\cite{svetnik2003random}, and they trained it to predict binding affinities using about 1200 protein--ligand complexes extracted from the PDBBind databases~\cite{wang2004pdbbind,wang2005pdbbind}. 
They reported a promising Pearson correlation coefficient $\rp=0.776$, which indicates a strong correlation between predicted and experimental binding affinities~\cite{liu2017forging}. Notably, the ML-based SF outperformed all considered classic ones. Later, a similar ML-based SF was specialized to virtual screening tasks, thus also addressing the comments of Ref.~\cite{gabel2014beware} on the poor performance of ML-based SFs in this task. 
Various subsequent studies have adopted more advanced ML models, such as dense neural networks (NNs)~\cite{zhu2020binding}, convolutional NNs~\cite{jimenez2018k, gomes2017atomic, seo2021binding, stepniewska2018development}, and graph NNs~\cite{li2021structure}, often further increasing the performance in binding affinity prediction. 
For a recent review on neural networks and deep learning techniques, see, e.g., Ref.~\cite{10.3389/frai.2020.00004}.
Different complex representations have been explored, ranging from distance-counts of protein--ligand atomic pairs (combined with kernel ridge regression)~\cite{Ballester2010,wojcikowski2017performance} to three-dimensional (3D) grids of atomic features (combined with convolutional NNs)~\cite{gomes2017atomic,jimenez2018k,stepniewska2018development,seo2021binding}. However, very recent studies have put these optimistic results into question~\cite{yang2020predicting}. They reported a drastic performance reduction when ML SFs are subjected to more stringent benchmarks. Noticeable examples are so-called vertical tests, whereby predictions are made for proteins not included in the training set, as opposed to the more common and less stringent horizontal tests, whereby a protein might be present both in the training and in the test sets, albeit bound to a different ligand. Along this line, Refs.~\cite{wojcikowski2017performance, gomes2017atomic} considered scaffolding tests designed to quantify how the prediction accuracy varies with the degree of structural similarity between the proteins in the training and in the test sets. Notably, Ref.~\cite{yang2020predicting} reported that the performance of an exemplary ML-based SF did not change when it was fed with only the protein structure, or only the ligand structure (as opposed to the whole complex). The authors thus suggested that the ML-based SF had not learned any information about the actual protein--ligand binding mechanism.

A critical problem for the development of ML techniques for computational docking is the limited number of experimental structures available for model training~\cite{warren2012essential,liu2017forging}. From the most relevant database, namely, the PDBBind repository, one may extract a few thousand complex structures, depending on the required resolution~\cite{warren2012essential}. While this number is steadily increasing, it is still orders of magnitude smaller than the amount of data typically used in other fields where ML has proven astonishingly successful, such as, e.g., computer vision. For example, state-of-the-art NNs for object detection are usually trained on databases including millions of images, e.g., the ImageNet database~\cite{5206848}. This issue is particularly relevant for deep NNs. These are often preferred to simpler ML regression models, such as kernel ridge regression or support vector machines, due to their superior generalization properties. However, in general, they require larger training databases to avoid overfitting problems. 
Beyond the small size, training databases might be affected by anthropogenic factors, leading to a biased selection towards particularly favourable instances, as recently found in studies on chemical reactions~\cite{2019Natur.573..251J}.
The contrasting findings discussed above call for further quantitative performance analyses on ML-based SFs, considering in particular the role of the training set and of the test type. One of the main questions we aim to address in this article is whether ML-based SFs can be trained using computer-generated complex structures created using docking software, providing access to larger and more tunable databases.

In this Article, we implement an SF using an NN with fully connected layers (FCNN). Following Refs.~\cite{Ballester2010,wojcikowski2017performance}, the descriptors used to represent the complex structures are counts of atomic pairs, one belonging to the protein and the other to the ligand, within various distance intervals. Two databases are used for training and testing. The first includes 2408 crystallographic structures extracted from the PDBBind database. Notably, we carefully check and prepare these structures with the inclusion of hydrogen atoms, in contrast to previous related studies that only considered heavy atoms. The second database includes 28,200 computer-generated structures created using the CCDC Gold docking engine within the MOE (Molecular Operating Environment) software interface~\cite{RefMOE,jones1997development,greenidge2016boosting}. 
In both databases, the complex structures are associated with the corresponding experimental binding affinities. The FCNN is trained to predict binding affinities of previously unseen complexes via a supervised learning algorithm. One of our goals is to quantify the performance of a FCNN combined with a distance-count description of protein--ligand complexes. Chiefly, we compare the performances reached using experimental and computer-generated databases, considering in both cases both horizontal as well as more challenging vertical tests. Finally, taking advantage of the creation of computer-generated databases, we explore the development of per-target SFs designed to predict binding affinities for a specific target protein. We consider 17 exemplary targets, training a specific SF for each of them using computer-generated databases including many complexes with different ligands docked into the same target protein. The performances obtained on the exemplary targets are analysed, also varying the size of the training set. In general, the obtained performances are encouraging, with noticeable variations depending on the target. To shed some light on these different performances, we make a comparison against basic linear regression models based on the molecular weight, optimized for each target.

\section{Materials and Methods} 

\subsection{Experimental Database} 
The first database includes 2408 complex structures obtained through X-ray crystallography~\cite{drenth2007principles}
and deposited into the Protein Data Bank~\cite{RefPDB,RefPDBurl}
. 
With this experimental technique, the positions of the heavy atoms, i.e., excluding hydrogen atoms, are estimated with a finite resolution.
We select structures with a resolution degree lower than 3 Å and whose ligand--target interaction information is available from the PDBbind database~\cite{wang2004pdbbind,wang2005pdbbind}. 
The 3D structures are manually checked for possible inconsistencies and curated with the addition of the hydrogen atoms, using the MOE software.  
This process is time consuming, thus limiting the size of the experimental database.
However, it allows us to adopt a more complete representation, as opposed to various previous related studies that considered only heavy atoms.
Furthermore, the inclusion of hydrogen atoms allows for performing a more direct comparison against the computer-generated complexes described below.
The target--ligand structures are associated with the corresponding
dissociation constant $K_d$. In fact, we consider the values of $pK_d \equiv-\log_{10} \left(K_d \right)$. 
These values are retrieved from the PDBbind database. For completeness, we report here the formal definition of dissociation~constant:
\begin{equation}
pK_d = -\log_{10} \left(K_d \right) = -\log_{10} \left(\frac{[P][L]}{[C]}\right),
\end{equation}
where $[P]$, $[L]$, and $[C]$ represent the concentrations of the protein, of the ligand, and of the complex, respectively.
Some relevant details on our databases are summarized in Table~\ref{table1}.

\begin{table}[H]
    \caption{Brief description of the experimental and the computer-generated databases. (*) The sign of the docking score is changed from negative to positive for consistency with the affinity values. These databases are freely accessible using the link in Ref.~\cite{pellicani_francesco_2022_7383354}.}
    \label{table1}

\begin{tabular}{|l|l|l|l|}
	\hline
	Database            & Number of complexes     & Mean affinity  & Mean docking score ($^*$)        \\
	\hline
	Experimental         & 2408     & 5.98 ($pK_d$)  &          \\
	\hline
	Computer generated          & 28,200     & 7.48 ($pK_i$) & 11.43         \\
	\hline
\end{tabular}

\end{table}

\subsection{Computer-Generated Database} 
The second database includes structures generated via computer simulations. We refer to it as the computer-generated database. To build it, the 3D structures of 17 selected target proteins are retrieved from the PDB repository. 
The selection focuses on diverse targets with many ligands deposited in the BindingDB database~\cite{chen2001bindingdb,chen2002binding,liu2007bindingdb}. 
Only complexes with a single ligand at the binding site and no cofactors are considered.  
For each target, a numerous set of ligands is retrieved. 
The choice of ligands is restricted among those whose experimental binding affinity for each selected target is available as expressed by the $pK_i$ score. This score measures the target--ligand affinity using a reference radio-ligand. Its value is generally close to $pK_d$. 
These ligands are docked into the respective target using the GOLD docking engine through the MOE software. 
Our simulations produce ten poses for each target--ligand pair using the GOLD docking engine. 
These poses are then rescored within MOE by the GBVI/WSA dG scoring function, following the protocols of previous studies; see., e.g., Refs.~\cite{FALSINI2019380,CENI2020112478}.
The target--ligand pose assigned of the best docking score is selected.
The final computer-generated database includes only the poses corresponding to the best score for each ligand at the respective target, totalling 28,200 protein--ligand complexes. 
It is worth pointing out that, due to the possible inaccuracy of the classical SF, the best pose does not necessarily match the correct experimental one. Still, avoiding mediocre scores likely increases the chances of excluding odd poses featuring clear inconsistencies, which might prevent the NN from learning to predict the correct affinity.
The number of pairs per target ranges from 384 for the PIM2 protein, to 6568 for the D2 protein. The selected target proteins and the corresponding number of complexes are summarized in Table~\ref{table2}. 
Evidently, the computer-generated database is significantly larger than the experimental one. This allows us to better analyse the learning speed of NNs. However, it is worth emphasizing that the docking pose created by the docking software is affected by the possible inaccuracy of the chosen docking engine. Instead, the 3D crystallographic structures are expected to correspond to the actual spatial configuration. 
It is worth pointing out that, in fact, spurious distortions might be present also in the crystallographic structures, but this is believed to rarely happen.
Clearly, while the binding affinity information present in the computer-generated database is, in fact, experimental, the 3D complex structure is affected by the choice of the SF used by the docking engine and, henceforth, by its possible inaccuracies.
It should be noted, however, that this does not necessarily represent a drawback. Indeed, in virtual screening campaigns, SFs are often used to select promising ligands from software-generated poses. Therefore, training the SF on the type of structures it will be asked to rank might actually be instrumental.

\begin{table}[H]
\caption{Breakdown of the protein--ligand complexes included in the computer-generated database.
The following legend defines the protein acronyms: 5HT2A: human {5-HT$_{2\mathrm{A}}$} receptor (pdbcode: 6A94); A2A: human {A$_{2\mathrm{A}}$} Adenosine receptor (pdbcode: 5NM4); ACE: Human acetylcholinesterase (pdbcode: 4EY5); BACE1: human BACE-1 enzyme (pdbcode: 6UVP); D2: human D2 dopamine receptor (pdbcode: 6CM4); DOP: human delta opioid receptor (pdbcode: 4N6H); FAAH: humanized variant of fatty acid amide hydrolase (pdbcode: 3PPM); GR: human glucocorticoid receptor (pdbcode: 4UDD); H1: human histamine {H$_{1}$} receptor (pdbcode: 3RZE); JAK1: human Janus kinase 1 (pdbcode: 6N7A); JAK2: human Janus kinase 2 (pdbcode: 6VN8); KOP: human kappa opioid receptor (pdbcode: 4DJH); M1: human M1 muscarinic acetylcholine receptor (pdbcode: 5CXV); MCL1: human Mcl-1 (pdbcode: 6UDV); OX2: human orexin 2 receptor (pdbcode: 5QWC); PI3K: human Phosphatidylinositol 4,5-bisphosphate 3-kinase catalytic subunit alpha isoform (pdbcode: 6PYS); PIM2: human PIM2 kinase (pdbcode: 4X7Q).
}
\label{table2}

\begin{tabular}{|l|l|l|l|l|l|l|l|l|l|}
	\hline
	Protein & 5HT2A & A2A & BACE1 & DOP & FAAH & GR & H1 & JAK1 & PI3K  \\          
	\hline
	N. of complexes & 2763 & 2914 & 1413 & 1243 & 508 & 843 & 1070 & 1213 & 1064           \\
	\hline
	Protein & PIM2 & ACE & KOP & M1 & MCL1 & JAK2 & OX2 & D2 &  \\
	\hline
	N. of complexes & 384 & 488 & 2431 & 1056 & 688 & 1394 & 2160 & 6568 &  \\
	\hline
\end{tabular}

\end{table}

\subsection{Complex Representation} 
\label{secrepr}
The input provided to the FCNN must be designed to represent, with good approximation, the 3D structure of the protein--ligand complexes. 
The description we adopt follows the archetypes proposed in previous studies~\cite{Ballester2010,wojcikowski2017performance}. 
Specifically, each descriptor corresponds to the count of atom pairs within a specified distance interval, whereby the first atom belongs to the target, the second to the ligand. 
We consider the following 10 atomic species: H, C, N, O, F, P, S, Cl, Br, and I.
This choice follows the study of Ref.~\cite{Ballester2010} (apart for hydrogen). 
See also the following studies of Refs.~\cite{wojcikowski2017performance,stepniewska2018development,seo2021binding}.
For the ligands, only the species H, C, N, O, P, and S are considered here.
This restriction is due to our choice of addressing only ligands not including halogens and metallic atoms.
While this is a possible limitation of our analysis, the selected atomic species are sufficient to describe many drug molecules and, therefore, we expect that our findings concerning training and testing protocols and per-target SFs are sufficiently general.
The above choices lead to 60 descriptors per distance interval. 
Notice that halogens could be avoided also for proteins. However, NNs quickly learn to ignore constant descriptors.
%
%
%
%
%
Notably, our representation takes into account H atoms. 
%
Furthermore, originally, Ref.~\cite{Ballester2010} considered only one distance interval, namely, 0--12 Å. Subsequently, Ref.~\cite{wojcikowski2017performance} adopted a more detailed representation, dividing the 12 Å range into six intervals. In this article, we explore different representations, varying both the number of intervals and their width. The goal is to identify the optimal compromise between representation accuracy and conciseness.
We point out that the original training matrix includes descriptors which might differ by orders of magnitudes. For this reason, a normalization operation is helpful. Our analysis on the experimental database shows that the most effective operation is dividing by the maximum value of all descriptors. 
To formally define this normalization procedure and, more in general, the descriptor vector, it is useful to introduce the following notation: the (unnormalized) number of pairs of atomic species $A\in(\mathrm{H, C, N, O, F, P, S, Cl, Br, I})$ (for the target) and $A^\prime\in(\mathrm{H, C, N, O, P, S})$ (for the ligand) that lie in the distance interval labeled by the index $k=1,2,\dots,k_{\mathrm{max}}$, where $k_{\mathrm{max}}$ is the chosen number of intervals, is denoted as $N_k^{AA^\prime}$. Therefore, we have 60 pairs of atomic species, meaning that each complex is characterized by $60k_{\mathrm{max}}$ descriptors. The interval widths we consider are $\ell=1.5$ Å, $\ell=2$ Å, or $\ell=3$ Å.
The intervals are defined by the following minimum and maximum distances: $r_{\mathrm{min}}=(k-1)\ell$ and $r_{\mathrm{max}}=k\ell$.
Different numbers of intervals $k_{\mathrm{max}}$ are addressed. The smallest value is $k_{\mathrm{max}}=1$ for all three interval widths. Instead, the largest number differs depending on the interval width, namely,  $k_{\mathrm{max}}=6$ for $\ell=1.5$ Å, $k_{\mathrm{max}}=7$ for $\ell=2$ Å, and $k_{\mathrm{max}}=5$ for $\ell=3$ Å. 
The normalized descriptors are $N_k^{AA^\prime}/\mathrm{max}\{N_k^{AA^\prime}\}$, where the maximum value is taken over all descriptors of all instances in the corresponding database.
This normalization is adopted for all results reported below.
%

%
%

\subsection{Target Values}
SFs are designed to predict a score proportional to the binding affinity. As already mentioned, we train and test our \SF using, as regression target, the $pK_d$ and the $pK_i$ 
values, for the experimental and the computer-generated databases, respectively. For the latter database, in only one analysis aiming at verifying if and to what extent a ML-based SF can mimic a classic SF, we also consider the docking score. For convenience, we adopt as a target value the negative of the docking score, so that higher values correspond to putatively higher affinities.
%
The mean target values of our databases are summarized in Table~\ref{table1}.

Notice that, during training, we adopt the standardized targets $d^{\prime} = \frac{d-\mu}{\sigma}$, where $d$ is the original value of binding affinity 
($pK_d$ or $pK_i$, depending on the considered database) or negative docking score (only when the \SF is trained to reproduce the prediction of the classic SF),
$\mu$ the database mean, and $\sigma$ the corresponding standard deviation.
This (linear) standardization does not affect the correlation coefficient $\rp$, and it is introduced only to allow us to better compare the phenomenology of the training processes on different databases. Furthermore, to favor comparison with other studies, the mean squared error (MSE) values reported below are computed considering un-normalized predictions and targets $d$, obtained by inverting the standardization formula.


\subsection{Regression Model and Training Protocol} 
\label{sectraining}
The goal of supervised learning is to train a regression model to map the complex descriptors to the target value, namely, the binding affinity (or the negative docking score).
The regression model adopted in this article is an FCNN, namely, a dense NN with all-to-all interlayer connectivity. 
The numbers of (hidden) layers $N_l$ and of hidden neurons per layer $N_h$ are chosen through the analysis described in Section~\ref{secresults}.
Note that $N_l$ does not count the descriptor layer, nor the output layer featuring a single neuron.
The activation function in the hidden layers is the hyperbolic tangent. 
The network weights and biases are optimized by minimizing the loss function, namely, the MSE between the network's predictions and the ground-truth target values.
To contrast possible overfitting phenomena, the loss function is augmented with a standard $L_2$ regularization term~\cite{goodfellow2016deep}. However, this does not lead to significant benefits, and the results we report hereafter correspond to a negligibly small regularization parameter. 
Our neural networks are implemented and trained using a very popular framework for deep learning, namely, the Keras Python library~\cite{chollet2015keras}.
Furthermore, we provide the code with the trained model corresponding to one of our most relevant SFs (see Section~\ref{secvertical}) at the repository of Ref.~\cite{pellicani_francesco_2022_7383354}.
The optimization is performed using a variant of stochastic gradient descent, named ADAM~\cite{https://doi.org/10.48550/arxiv.1412.6980}.
An adequate mini-batch size turns out to be around 50 and 200, for the experimental and the computer-generated databases, respectively.

The training epochs are iterated until the prediction accuracy on the test set stops improving, i.e., before entering the regime where overfitting phenomena dominate. Due to the small size of the experimental database, introducing a validation set is not practical. For consistency, the same protocol is adopted also for the computer-generated database. This is intended to estimate the optimal potential performance. While, in principle, it might lead to a slight overestimation, this effect is found not to be important. In particular, in the most critical per-target \SF tests, the prediction accuracy is found to be quite stable as a function of the training epochs.
%
To monitor the prediction accuracy, two metrics are considered, namely, the Pearson correlation coefficient $\rp$ and the MSE. These are computed on a test set which includes around 20\% of the whole database, while the remaining instances are used for training. %
The training and testing process is repeated ten times, considering just as many different random splittings between training and test instances, or simply different (random) selections of training data and mini-batches for gradient descent. The accuracy scores reported hereafter correspond to the average of the test scores, while error bars correspond to the estimated standard deviation of the average.
%
This averaging procedure avoids spurious fluctuations due to accidentally favourable or adverse selections of the test instances, providing a more reliable estimate of the prediction accuracy in a realistic scenario.

\section{Results}

\label{secresults}


%
\subsection{Selection of Descriptors and Network Structure} 
\begin{figure}[H]
    \includegraphics[width=0.8\columnwidth]{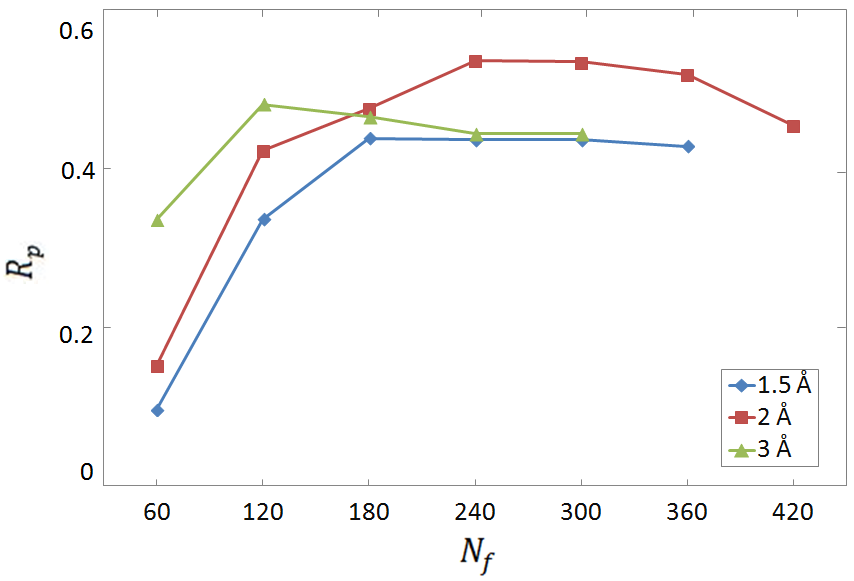}
    \caption{Pearson correlation coefficient $\rp$ for the predictions of binding affinity provided by the \SF, as a function of the number of descriptors $N_f$. This \SF is trained and tested on the experimental database. The protein--ligand complex structure is represented by atomic-pair counts within a variable number of distance intervals. The three datasets correspond to different interval widths. Sixty pairs of atomic species are considered, including pairs with hydrogen.}
    \label{fig1}
\end{figure}

To identify the optimal choice for the complex--structure representation, we analyse the prediction accuracy on test sets of 
300 randomly chosen experimental complexes, as a function of the number of distance intervals of atom-pair counts. 
The choice of considering 300 complexes (randomly selected from the whole experimental database) for testing is a trade off between the need of the largest possible training set and the minimal number required for a reliable test. As discussed in Section~\ref{sectraining}, 10 random non-overlapping splittings are considered to reduce the role of statistical fluctuations due to particularly favourable or adverse testing complexes.
%
Notice that, while the test protein--ligand complexes are distinct from those used for the optimization of weights and biases, some target proteins might occur in both training and test sets, albeit bound to a different ligand.
This is what we refer to as a horizontal test.
The results are shown in Figure~\ref{fig1}, considering the three interval widths, namely, $\ell=1.5$ Å, $\ell=2$ Å, and $\ell=3$ Å. 
For a complete definition of the descriptor vector, we refer the reader to Section~\ref{secrepr}.
One notices that the maximum $\rp$ score, which corresponds to the optimal representation, is obtained using $k_{\mathrm{max}}=4$ intervals of width $\ell=2$ Å, meaning that atom pairs are counted only if the two atoms are less than 8 Å apart. Considering that 60 pairs of atom species are considered, the total number of descriptors is 240. 
This representation is adopted for all results reported~hereafter.

\begin{figure}[H]
    \includegraphics[width=0.8\columnwidth]{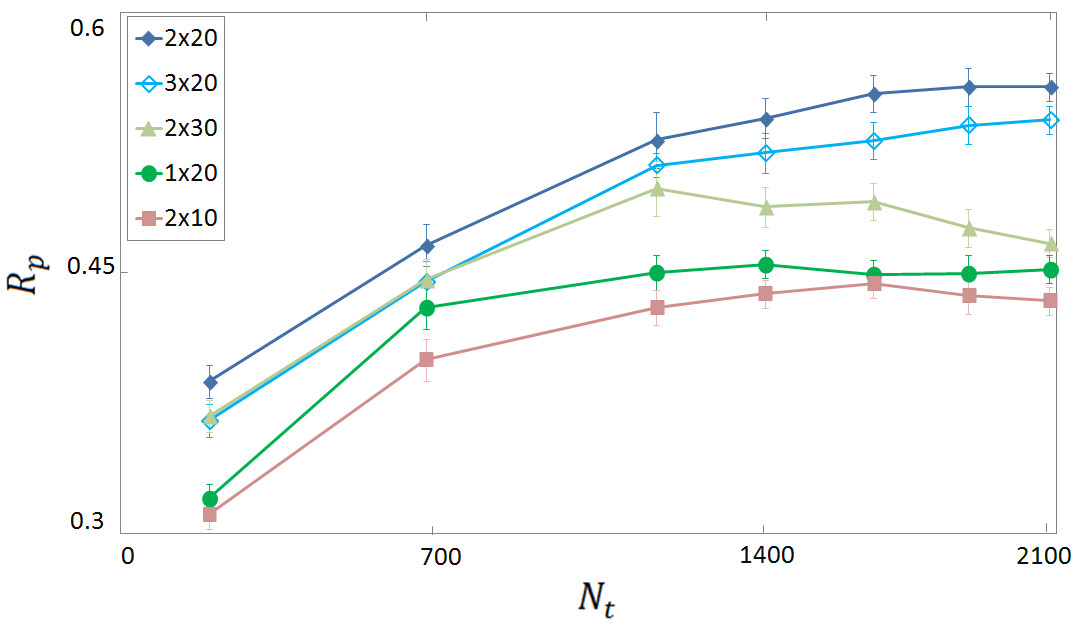}
    \caption{$\rp$ score for binding affinity prediction as a function of the size $N_t$ of the (experimental) training set. Each dataset corresponds to a choice $N_l \times N_h$ for the number of hidden layers $N_l$ and of neurons per layer $N_h$ in the fully connected neural network (FCNN). The complex representation includes $N_d=60$ descriptors, with four distance intervals of width 2A.
    }
    \label{fig2}
\end{figure}

We identify the optimal depth and width of the neural network by analysing the $\rp$ score on the experimental test set. The results are shown in Figure~\ref{fig2} as a function of the number of training instances  $N_t$. Different numbers of layers $N_l$ and of neurons per (hidden) layer $N_h$ are considered.
The highest performance is obtained for $N_l\times N_h=2\times 20$.
An analogous analysis performed on the computer-generated data (not shown) indicates that, in that case, the optimal network structure is $N_l\times N_h=4\times 40$. The need of a deeper network can be attributed to the larger size of the computer-generated database. Indeed, it is known that deeper NNs are more effective in extracting useful information from larger databases, while they are more susceptible to overfitting phenomena when the database is sparse.
These two NN structures are adopted for all results reported below, unless otherwise specified.

\subsection{Horizontal Tests on Experimental and Computer-Generated Structures}
\label{horizontal}
\begin{figure}[H]
    \includegraphics[width=0.8\columnwidth]{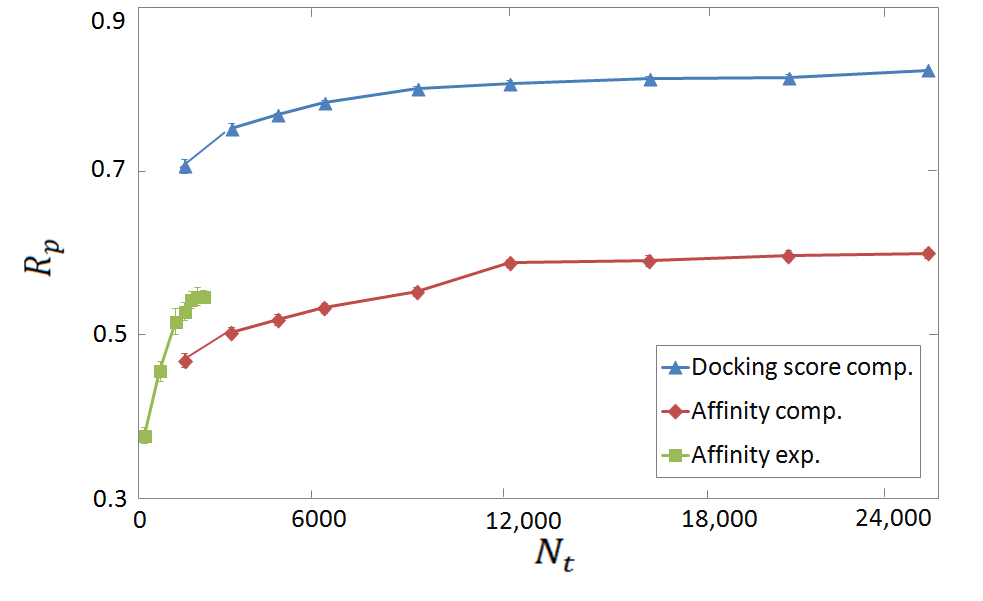}
    \caption{{$\rp$ prediction--accuracy} 
 score as a function of the training set size $N_t$. The green squares and the red rhombi correspond to the binding affinity prediction by the \SF trained on the experimental complexes  and on the computer-generated databases, respectively. The blue triangles correspond to the predictions of the docking score by the \SF trained on the computer-generated database. The network structure is described in Figures~\ref{fig1} and~\ref{fig2}.}
    \label{fig3}
\end{figure}

\begin{figure}[H]
    \includegraphics[width=0.8\columnwidth]{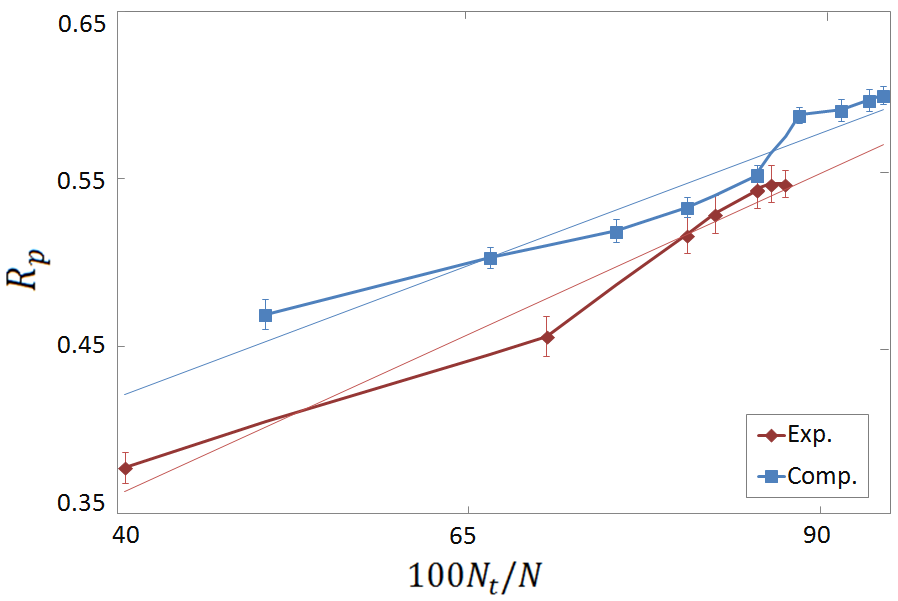}
    \caption{$\rp$ prediction--accuracy score as a function of the percentage  of the training set size $N_t$ compared to the whole database size $N$, namely, $100N_t/N$. The red rhombi and the blue squares correspond to the \SF trained on the experimental complexes  and on the computer-generated complexes, respectively.}
    \label{fig4}
\end{figure}

One of our main goals is to compare the performances of ML-based SFs trained and tested on experimental and on computer-generated databases.
The two learning curves for our \SF are compared in Figure~\ref{fig3}, where the $\rp$ score for the binding affinity prediction is plotted as a function of the training set size $N_t$.
As a term of comparison, the $\rp$ value obtained when the (negative) docking score is used as target value, both during training and testing phases, is also shown.
Noticeably, remarkably high performances are obtained in this latter test, namely, $\rp\simeq 0.82$.
This indicates that the chosen combination of representation and regression models is capable of learning, within good approximation, the function corresponding to the docking score.
Clearly, while this result suggests that a ML-based SF can at least imitate a classic SF, it does not imply a good performance for binding affinity prediction since, as already discussed, classic SFs do not always perform well in this task.
Indeed, the maximum $\rp$ score corresponding to the binding affinity prediction is somewhat smaller, namely, $\rp \simeq 0.55$ and $\rp \simeq 0.60$, for the experimental and the computer-generated databases, respectively. These scores still correspond to moderately strong correlations between predictions and ground-truth (i.e., experimental) binding affinities.
The lower score at the maximum $N_t$ on the experimental database can be attributed to the larger size of the computer-generated one. Actually, it appears that the learning is faster on the experimental database, since higher accuracies are in fact obtained when the training set size $N_t$ is comparable.
To further inspect this effect, the same $\rp$ scores are plotted in Figure~\ref{fig4} as a function of the percentage of training instances compared to the size $N$ of the whole database.
One observes an approximately linear increase. The slopes corresponding to the experimental and to the computer-generated databases are comparable. This suggests that the performance improves due to the increasing probability of finding the same or similar proteins in both the training and the test sets. In concord with this supposition, the $\rp$ score is consistently higher on the computer-generated database, whereby the number of proteins is lower.
A similar supposition has been put forward in Ref.~\cite{yang2020predicting}, and it is supported also by the vertical tests discussed below.

\unskip

\subsection{Vertical Tests}
\label{secvertical}
\begin{figure}[H]
    \includegraphics[width=0.8\columnwidth]{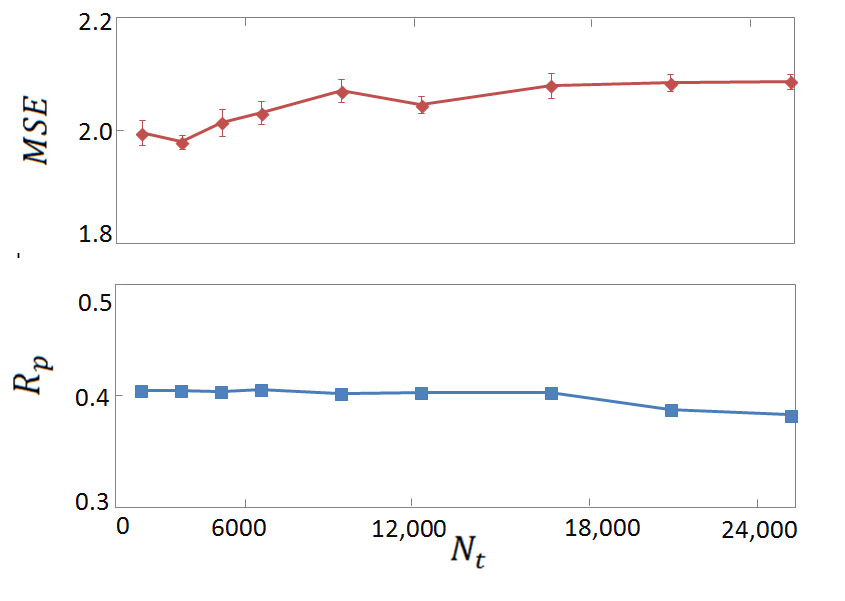}
    \caption{{Mean} 
 squared error (MSE, upper panel) and $\rp$ score (lower panel) for binding affinity prediction in the vertical test, as a function of the training set size $N_t$. The \SF   is trained on a computer-generated database including complexes made from 13 proteins, and tested on complexes made from four proteins not included in the training set.}
    \label{fig5}
\end{figure}

\begin{figure}[H]
    \includegraphics[width=0.8\columnwidth]{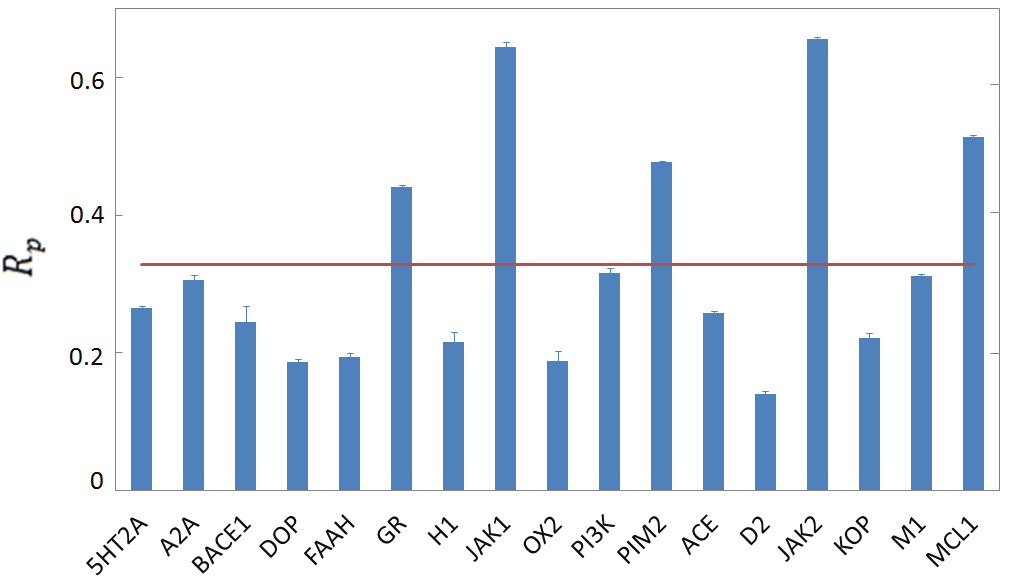}
    \caption{{Accuracy scores} 
 $\rp$ for binding affinity affinity predictions for 17 \SF's in per-target vertical tests. Each SF is trained on the computer-generated database excluding the complexes made from the protein indicated on the horizontal axis, and tested on the excluded complexes. The horizontal red line indicates the average score.}
    \label{fig6}
\end{figure}

The horizontal tests described above provide encouraging results. As anticipated, however, these might be over-optimistic, being biased by the similarities among the complexes present in the training and in the test sets.
In a real-case scenario, universal SFs are expected to describe the binding strength of candidate ligands into novel proteins under investigation. Quite likely, these proteins are dissimilar from those included in the databases available at the time of model definition.
A fairer performance assessment is therefore provided by so-called vertical tests, whereby  complexes made from proteins present in the test set are excluded from the training set.
The first vertical test we consider is performed on computer-generated complexes made from the four proteins FAAH, PIM2, ACE, and MCL1, totaling 2068 complexes.
The \SF  is trained on $N_t$ complexes made from the remaining 13 proteins of our computer-generated database (see Table~\ref{table2}).
The $\rp$ score is shown in Figure~\ref{fig5} as a function of $N_t$. One notices a significantly reduced performance compared to the horizontal test. The accuracy score, $\rp \approx 0.4$, corresponds to an only moderate correlation between predicted and experimental binding affinities. 
It is worth mentioning that a similar vertical test on computer-generated complex structures was performed also in Ref.~\cite{wojcikowski2017performance} using a random-forest regression model. That study reports an even lower score, namely, $\rp \simeq 0.2$. We attribute the better performance of our \SF to the use of an FCNN, which is more suited for extracting useful information from large databases.
It is also worth comparing the performance of our \SF with the one displayed by a popular classic SF. We consider the GBVI/WSA dG scoring function of the MOE software.
Since the corresponding docking score is negative, with lower values corresponding to more favourable poses, we consider the negative of the docking score. Its correlation with the binding affinity turns out to be only marginal, corresponding to $\rp \simeq 0.2$. 
To favor further comparative studies, the code with our \SF, trained on the whole subset including the 13 groups of complexes discussed above, is provided via the repository of Ref.~\cite{pellicani_francesco_2022_7383354}. 
While our \SF seems to perform relatively better than the two considered benchmarks, its performance is not fully satisfactory. 
Chiefly, one notices that the performance does not improve with the training set size $N_t$.
To further elucidate this finding, we perform a series of per-target vertical tests. Seventeen \SF's are trained on just as many computer-generated databases obtained by excluding all complexes made from each of the 17 proteins. The excluded complexes are used as test sets for the corresponding SF.
The 17 corresponding $\rp$ scores are shown in Figure~\ref{fig6}. Again, the performance is, on average, appreciably lower compared to the horizontal test discussed above.
This corroborates the claim that horizontal tests provide over-optimistic performance measures, probably due to the correlations and similarities among training and test complexes.
In addition, the large performance fluctuations among the 17 \SF's corresponding to the different targets are noteworthy.
For example, remarkably high scores are obtained for the JAX1 and JAX2 proteins. These might be attributed to the similarity between these two proteins, meaning that including in the training set complexes derived from one of them allows the \SF   learning how to predict binding affinities for the other. However, the close relationship between binding affinity and the ligand molecular weight might also play a role. This is further discussed below.

\unskip

\subsection{Per-Target Scoring Functions}
\begin{figure}[H]
    \includegraphics[width=0.8\columnwidth]{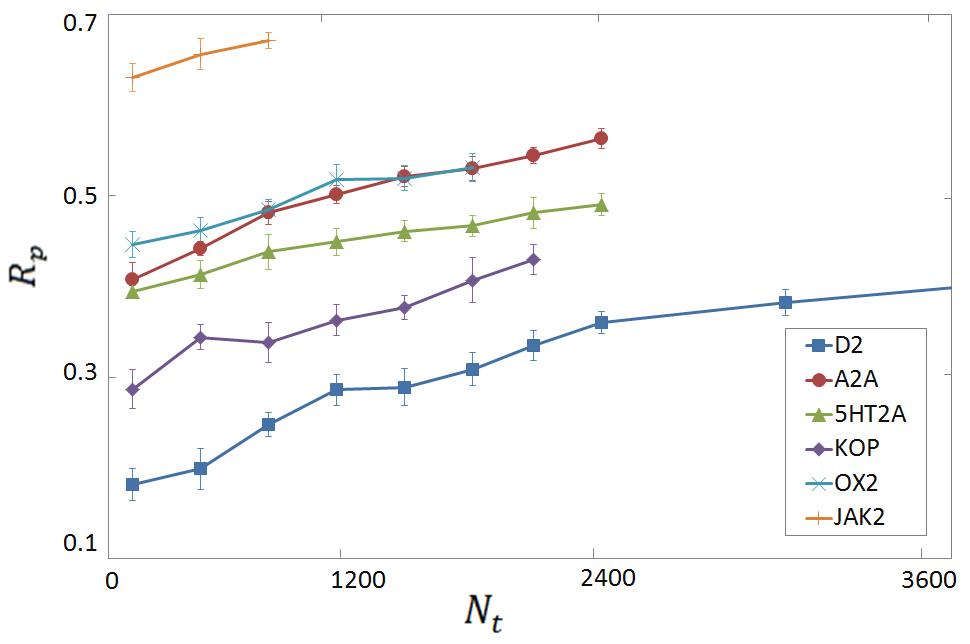}
    \caption{$\rp$ score for binding affinity predictions from six per-target \SF's, as a function of the training set size $N_t$. Each SF is trained and tested on computer-generated complexes made from the protein indicated in the legend.}
    \label{fig7}
\end{figure}

\begin{figure}[H]
    \includegraphics[width=0.8\columnwidth]{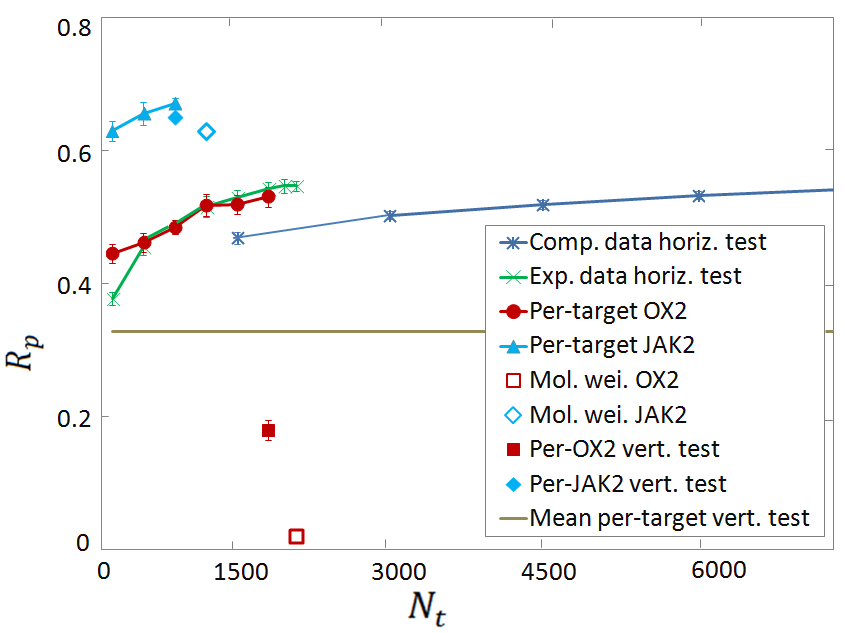}
    \caption{$\rp$ accuracy scores as a function of the training set size $N_t$, for two exemplary target proteins, namely, JAK2 (cyan) and OX2 (red). 
    The performances of the per-target \SF's (cyan triangles for JAK2 and red circles for OX2) are compared to the corresponding scores of the \SF in the vertical test (cyan full diamond and red full square) and of the molecular weight linear regression (cyan empty diamond and red empty square). The large green x's correspond to the universal \SF in the horizontal test on the experimental database, while the small blue x's  correspond to the analogous test of the computer-generated database. The horizontal gray line represents the average score on the 17 per-target vertical tests.}
    \label{fig8}
\end{figure}

The unremarkable performances displayed by the universal ML-based SF in the vertical tests lead us to explore different strategies.
As discussed in Section~\ref{horizontal}, the experimental and the computer-generated databases appear to be comparably effective for training SFs for binding affinity predictions.
Clearly, the computer-generated complexes can be relatively easily generated. This suggests the idea of developing per-target SFs on-demand, whenever a novel protein is targeted.
Such a SF would be trained on a purposely created database including only complexes made from the target protein.
To explore this direction, we consider the six proteins with more complexes in our computer-generated database, namely, D2, A2A, 5HT2A, KOP, OX2, and JAX2.
Six per-target \SF's are trained and tested only on complexes made from one protein. 
%
Motivated by the sizes of the six databases, we adopt FCNNs with $N_l \times N_h=2 \times 20$, apart for the D2 target, for which the parameters $N_l \times N_h = 3 \times 20$ are expected to be more adequate.
The corresponding $\rp$ scores for binding affinity prediction are shown in Figure~\ref{fig7}, as a function of the training set size $N_t$. The tests are performed on 300 complexes.
%
The performances are relatively high. The average $\rp$ obtained using, for each target, 
$N_t=900$ training instances, is $\rp =  0.44$. When the largest available $N_t$ for each target is employed, the average score is $\rp = 0.52$. 
These scores are to be compared with the average vertical test on these six targets, corresponding to $\rp =  0.30$. Chiefly, in all six tests, the $\rp$ score systematically increases with $N_t$, suggesting that sufficiently performant per-target SFs can be obtained when adequately large computer-generated training databases are available.
There are also noticeable performance differences among the six targets, ranging from $\rp \approx 0.4$ for D2, to $\rp \approx 0.67$ for JAK2.
To shed some light on these differences, we compare the per-target \SF's to simple linear regression models. Specifically, we assume the linear law $pK_i= A+ m B$, where $m$ is the ligand molecular weight (MW), and $A$ and $B$ are the fitting coefficients. These are fixed via MSE minimization on the whole per-target database.
The comparison between the per-target \SF and the corresponding MW SF is shown in Figure~\ref{fig8}, for two exemplary targets, namely the JAK2 and OX2 proteins.
Notice that the corresponding scores in the per-target vertical tests are also shown.
One notices that, for the JAK2 protein, even the simple MW SF provides a remarkable performance, namely, $\rp \simeq 0.62$. 
Tentatively, this effect might be attributed to the large pocket size at the binding site for this particular protein, meaning that the binding strength is simply proportional to the ligand size. 
Clearly, additional analyses would be required to further corroborate this tentative explanation.
Assuming this explanation is indeed sound, it not so surprising that both the per-target \SF and the universal \SF in the per-target vertical test provide comparable (but still superior) scores.
Instead, for the OX2 protein, the predictions of the MW SF have essentially zero correlation with the binding affinity, corresponding to $\rp \approx 0$.
The per-target \SF reaches an encouraging $\rp \simeq 0.53$ at the largest $N_t$.
Notably, this per-target \SF significantly overcomes the score of the universal \SF in the corresponding per-target vertical test ($\rp \simeq 0.18$).
These findings suggest that per-target \SF's are able to learn non-trivial mappings from computationally feasible computer-generated databases.

\unskip

\section{Discussion}
We have analysed the performance of scoring functions (SFs) based on fully-connected neural networks (dubbed \SF's) in predicting the binding affinities of protein--ligand complexes from a representation of the 3D complex structure.
Our choice for the structure representation is based on atomic-pair counts within suitably chosen distance intervals, and we investigated the optimal number and width of such intervals.
The effectiveness of crystallographic 3D structures for SF training has been compared to that of computer-generated structures created with popular commercial docking-simulation software.
Notably, the two types of data turn out to be comparably effective, suggesting that more accessible computer-generated database can be used for SF development, rather than the less copious experimental structures.
Importantly, our \SF's have been assessed in horizontal tests as well as in vertical tests, whereby no protein is included both in the training and in the test databases.
A significant performance degradation is found in the latter, as in fact reported also in the study of Ref.~\cite{wojcikowski2017performance} using different regression models compared to the neural networks adopted here.
This corroborates the contention of the authors of Ref.~\cite{yang2020predicting}, who argued that SFs based on machine learning might not learn the ligand--target binding mechanism, but rather the correlations among complexes present in the training and in the test sets.
These findings indicate that vertical tests represent fairer assessments for the performances of SFs in real-case scenarios.
The relative easy of creating computer-generated databases led us to explore the development of per-target \SF's. Six exemplary targets have been considered, obtaining encouraging results. Chiefly, a systematic performance improvement with the training set size was observed.
Furthermore, we found instructive results by comparing the \SF's to linear-regression models based on the molecular weights. Interestingly, in some cases, even such simple theories reach high correlation with experimental binding affinities.

SFs are an important tool to accelerate drug discovery. Massive research endeavours have been devoted to implementing different families of SFs.
However, the development of SFs based on machine learning is still in an early stage. The variable and sometimes contrasting results reported in the literature indicate that more research is due to establish reliable benchmarking protocols. 
We argue that the ease of generating training data via computer simulations, together with the first encouraging findings reported in this article, will favour further research endeavours aiming at developing per-target SFs based on machine learning, tailored at specific proteins or protein families. 
Additional techniques borrowed from the field of artificial intelligence might help with facing the problems associated with sparse, possibly biased training databases~\cite{PMID:29680150}. Along this line, generative neural networks have been employed to implement de-novo molecular design, avoiding virtual screening of excessively large databases~\cite{doi:10.1021/acs.jcim.8b00839}.
Further research should also consider the use of additional complex descriptors, taking into account, e.g., entropic contributions, hydrogen-bond information, and/or pharmacophore models.
Extending the considered atomic species, addressing, e.g., halogens, is also important; indeed, halogens species are included in many drug-like molecules.
To facilitate future comparative studies, we provide our databases (experimental and computer-generated) via the link available in Ref.~\cite{pellicani_francesco_2022_7383354}, including the 3D structures, deposited in PDB files, as well as the binding affinities and the docking scores.
This repository also stores the code with one of our \SF's.
The provided SF is trained on a subset of our computer-generated database, as discussed in Section~\ref{secvertical}.

\vspace{6pt}

\acknowledgments{Fruitful discussions with {Pierbiagio Pieri and Andrea De Simone are acknowledged.
We also thank Andrea Spinaci} 
for support during data preparation.
{We also acknowledge 6 Tour S.r.l for the fruitful interactions in the initial phase of this work.}
This research was funded by the University of Camerino under the FAR2018 project titled ``Supervised machine learning for quantum
matter and computational docking'' and by the Italian MIUR under the project PRIN2017 CEnTraL 20172H2SC4.
S.P. acknowledges PRACE for awarding access to the Fenix Infrastructure resources at Cineca, which are partially funded by the European Union’s Horizon 2020 research and innovation program through the ICEI project under Grant Agreement No. 800858.
}








\bibliography{bibtexrefs}

\end{document}